\documentclass[aps,prd,longbibliography,secnumarabic,preprint,showpacs]{revtex4-1}
\usepackage{mathrsfs,amsmath,amssymb,bm,braket,here,comment,framed,ulem,cancel,ascmac}
\usepackage{etoolbox}
\usepackage{breqn}

\usepackage{color, graphicx}
\usepackage[colorlinks=true]{hyperref}
\usepackage[hang,small,bf]{caption}
\usepackage[subrefformat=parens]{subcaption}
\makeatletter
\let\cat@comma@active\@empty
\makeatother

\begin{document}

\preprint{KOBE-COSMO-19-17}
\title{Stability of Axion Dark Matter--Photon Conversion}%

\author{Emi Masaki}%
\email[]{emi.masaki@stu.kobe-u.ac.jp}

\author{Arata Aoki}%
\email[]{arata.aoki@research.phys-sci.com}

\author{Jiro Soda}%
\email[]{jiro@phys.sci.kobe-u.ac.jp}

\affiliation{Department of Physics, Kobe University, Kobe 657-8501, Japan}%
\date{\today}



\begin{abstract}
	It is known that a coherently oscillating axion field is a candidate of the dark matter. 
	In the presence of the oscillating axion, the photon can be resonantly produced through the parametric amplification. 
	In the universe, there also  exist cosmological magnetic fields which are coherent electromagnetic fields. 
	In the presence of magnetic fields, an axion can be converted into a photon, and {\it vice versa}.
	Thus, it is interesting to investigate what happens for the axion--photon system in the presence of 
	both the axion dark matter and the magnetic fields. 
	This system can be regarded as a coupled system of the axion and the photon
	whose equations contain the Mathieu type terms.
	We find that the instability condition is changed 
	in the presence of magnetic fields in contrast to the conventional Mathieu equation.
	The positions of bifurcation points between stable and unstable are shifted and new instability bands appear.
	This is because the resonantly amplified axion can be converted to photon, and {\it vice versa}.
\end{abstract}

\maketitle
\tableofcontents
\newpage
\section{\label{sec:intro}Introduction}
The cosmological dark matter problem has been studied in the context of beyond the standard model of particle physics.
One of such dark matter candidates is an axion which 
has been originally proposed as a solution for the strong $CP$ problem 
~\cite{10.1103/PhysRevLett.38.1440,10.1103/PhysRevD.16.1791,10.1103/PhysRevLett.40.223,10.1103/PhysRevLett.40.279}. 
This original axion is called a QCD axion.
String theory predicts axionlike particles (ALPs) 
with a broad mass range~\cite{10.1088/1126-6708/2006/06/051,10.1103/PhysRevD.81.123530}.
Throughout this paper, we will simply use a word “axion”, for both QCD axion and ALPs.
The axion has feeble interaction with standard model particles
and could be produced in the early universe by nonthermal mechanism.
This is the reason why the axion can be the dark matter
~\cite{10.1016/0370-2693(83)90637-8,10.1016/0370-2693(83)90638-X,10.1016/0370-2693(83)90639-1}.
In particular, it is known that 
ultralight axions called fuzzy dark matter~\cite{10.1103/PhysRevD.95.043541}
can resolve the issues in $\Lambda$CDM,
{\it e.g.}, the core-cusp problem and the missing satellite problem.
We can treat axion dark matter as a classical field.
Then, the axion is coherently oscillating with a frequency determined by the mass.
There are various experiments to search for the axion dark matter
~\cite{10.1103/PhysRevLett.121.261302,10.1103/PhysRevLett.122.121802}.

It is known that, in the presence of the axion dark matter, the propagation of photons 
is governed by the Mathieu equation~\cite{10.1093/ptep/pty029}.
The properties of the Mathieu equation are well studied in mathematics~\cite{Mathieu1868,978-0486612331,10.1115/1.4039144}.
It is known that the system becomes unstable for specific parameter regions.

In the universe, on top of the axion dark matter which is a coherent axion field, 
there exist cosmological magnetic fields which is a coherent electromagnetic field.
Remarkably, in the presence of magnetic field, there occurs the axion--photon conversion~\cite{10.1016/0370-2693(86)90869-5,10.1103/PhysRevD.37.1237}.
The axion--photon conversion has been investigated in the context of astrophysics.
Indeed, the axion--photon conversion can explain the fact that high energy photons
can reach the Earth through intergalactic magnetic fields without disappearing.
On the other hand, in the CAST experiment~\cite{10.1038/nphys4109}, 
strong magnetic field is applied to the detector in order to detect axions produced in the sun by converting axions into photons.
The fact that no signal of axions has been detected until now has given 
constraints on the mass of the axion and the coupling constant between an axion and and two photons.

As we explained in the above, it is natural to consider the axion dark matter and the magnetic fields
at the same time.
Hence, in this paper, we investigate what
happens for the axion--photon system in the presence of 
both the axion dark matter and the magnetic fields. 
More precisely, we study the stability of such system in terms of both numerical and analytical methods.
Although there are related papers which discuss behavior of axion dark matter and photon
with and without magnetic field
~\cite{10.1103/PhysRevLett.51.1415,
10.1103/PhysRevD.32.2988,
10.1103/PhysRevD.36.974,
0.1016/0370-2693(95)01393-8,
10.1134/S1063776109030030,
10.1103/PhysRevD.85.025010,
10.3204/DESY-PROC-2013-04/espriu_domenec,
10.1142/S0217751X15500992,
10.1103/PhysRevD.97.123001,
10.1103/PhysRevLett.121.241102,
10.1088/1475-7516/2018/11/004,
10.1140/epjc/s10052-019-6759-7},
to the best of our knowledge, no one did the stability analysis focusing on the
axion--photon conversion.

This paper is organized as follows.
In Sec.~\ref{sec:formulation}, we introduce basic equations of axion electrodynamics.
We derive basic equations by separating a background and perturbed quantities.
Then, we show numerical results in Sec.~\ref{sec:Ince--Strutt Chart}.
They show stability of the solutions for the basic equations.
In Sec.~\ref{sec:transition_curves}, we give an analytical derivation of the numerical results.
In particular, we will show you how to determine the boundaries
between stable and unstable region in the parameter space.
In Sec.~\ref{sec:diagram}, we will discuss an interpretation of our numerical results
and a possible application.
The final Sec.~\ref{sec:conclusion} is devoted to the conclusion.

\section{\label{sec:formulation} Axion Electrodynamics}
In this section, we introduce basic equations of axion electrodynamics.
Then, we consider an oscillating axion field and a static uniform magnetic field as a background.
Given the background, we derive perturbative equations for describing propagation of axions and photons. 
With these equations, we can study mixing between axions and the photons and the stability of the system.

\subsection{\label{subsec:Axion Electrodynamics}Basic Equations of Axion Electrodynamics}
We consider the following system:
\begin{align}\label{eq:action}
	S = \int d^4x \left[-\,\frac{1}{2}\left(\partial_\mu\,a\,\partial^\mu\,a + m^2_a\,a^2\right)
	-\frac{1}{4}\,F_{\mu\nu}F^{\mu\nu}-\frac{1}{4}g_{a\gamma\gamma}\,aF_{\mu\nu}\tilde{F}^{\mu\nu}\right]\ ,
\end{align}
where $a$ is an axion field with mass $m_a$, and $g_{a\gamma\gamma}$ is a coupling constant.
The field strength $F_{\mu\nu}$ of the electromagnetic field $A_\mu(\vec{x}, t)$
and its dual $\tilde{F}_{\mu\nu}$ are given by
\begin{align}
	F_{\mu\nu}\equiv \partial_\mu A_\nu-\partial_\nu A_\mu \ ,\quad
	\tilde{F}_{\mu\nu}\equiv\frac{1}{2}\epsilon_{\mu\nu\rho\sigma}F^{\rho\sigma}.
\end{align}
Using the potential $A_\mu(\vec{x}, t) 
= \left[-\phi(\vec{x}, t) ,\vec{A}(\vec{x}, t)\right]$, the electric and magnetic fields are defined by
\begin{align}
	&\vec{E}(\vec{x}, t)=-\partial_{t}{\vec{A}(\vec{x}, t)} - \nabla \phi(\vec{x}, t) \ ,\\
	&\vec{B} (\vec{x}, t)
	= \nabla \times \vec{A}(\vec{x}, t)\ .
\end{align}
We can get the equations for the axion
\begin{align}\label{eq:Klein-gordon}
	&(\Box-m^2_a)\,a(\vec{x}, t)=g_{a\gamma\gamma}\,\vec{E}(\vec{x}, t)\cdot\vec{B}(\vec{x}, t)\ ,
\end{align}
and for electromagnetic fields
\begin{align}\label{eq:Maxwell}
	\begin{cases}
		\Box\,\phi(\vec{x}, t)=-g_{a\gamma\gamma}\,\vec{B}(\vec{x}, t)\cdot\left[\nabla a(\vec{x}, t)\right]\ ,\\
		\Box\,\vec{A}(\vec{x}, t)=g_{a\gamma\gamma}\,\left[\left[\partial_t\,{a}(\vec{x}, t)\right]\,\vec{B}(\vec{x}, t)
		+\left[\nabla\,a(\vec{x}, t)\right]\times\vec{E}(\vec{x}, t)\right]\ .
	\end{cases}
\end{align}
Here, we have chosen the Lorenz gauge:
\begin{align}\label{eq:Lorenz gauge}
	\nabla \cdot \vec{A}(\vec{x}, t) +\partial_t\, \phi(\vec{x}, t) = 0\ .
\end{align}
Equations (\ref{eq:Klein-gordon}) and (\ref{eq:Maxwell}) are basic equations of axion electrodynamics~\cite{10.1103/PhysRevLett.58.1799}.
For the full analysis, we need to resort to lattice calculations. Here, we use the perturbative analysis.

\subsection{\label{subsec:Background_Equations}Background Equations}
Now, we assume both the axion dark matter and the magnetic fields as a background.
The background magnetic field is static and uniform,
\begin{align}
	\vec{B}_0=[B_0,0,0] \ .
\end{align}
We introduce here coordinate basis so that the propagation is in the direction $\vec{e}_z=[0,0,1]$,
one of the rests is parallel to the magnetic field: 
$\vec{e}_\parallel=[1,0,0]$, and the other is $\vec{e}_\perp=[0,1,0]$.
The background equation for axion is 
\begin{align}\label{eq:axion_zeromode}
	\partial^2_t\,{a}_0(t)+m^2_a\,a_0(t)= -g_{a\gamma\gamma}\,B_0\,E_{0\parallel}(t)\ ,
\end{align}
and for photon 
\begin{align}\label{eq:photon_zeromode}
	\partial_t\,E_{0\parallel}(t) = g_{a\gamma\gamma}\,B_0\,[\partial_t\,{a}_0(t)]\ .
\end{align}
Note that we have chosen the radiation gauge
\begin{align}
	 \phi(t) = 0\ ,\ \ \nabla \cdot \vec{A}(\vec{x}) =0\ .
\end{align}
This is because the source term of the scalar potential $\phi(t)$ equation vanish
\begin{align}
	\Box\,\phi(t) = 0\ .
\end{align}
Solving the Eq.~(\ref{eq:photon_zeromode}), 
we see that the electric field is induced by the axion oscillation:
\begin{align}\label{eq:induced_electric_filed}
	E_{0\parallel}(t) = g_{a\gamma\gamma}\,B_0\,{a}_0(t)\ .
\end{align}
Substituting (\ref{eq:induced_electric_filed}) into (\ref{eq:axion_zeromode}),
we can get 
\begin{align}
	\ddot{a}_0(\tau) + \Omega^2_\beta\, a_0(\tau)=0\ ,
\end{align}
where we replace time variable $t$ with $\tau \equiv m_at$, and express a derivative with respect to $\tau$ by dot.
Here we also introduced new dimensionless parameters $\beta$ and $\Omega_\beta$ as
\begin{align}
	&\beta \equiv \frac{g_{a\gamma\gamma}\,B_0}{m_a}
	= 1.95\times10^{-9}\left(\frac{10^{-22}\,{\rm eV}}{m_a}\right)
	\left(\frac{g_{a\gamma\gamma}}{10^{-11}\,{\rm GeV}^{-1}}\right)\left(\frac{B_0}{10^{-9}\,{\rm G}}\right)\ ,\\
	&\Omega_\beta\equiv\sqrt{1+\beta^2}\ .
\end{align}
Note that conservatively we have a constraint 
$g_{a\gamma\gamma} \leq 10^{-11} {\rm GeV}^{-1} $.
Recall the relation $1 {\rm G} = 1.95 \times10^{-2} \ {\rm eV}^{2}$,
for the cosmological magnetic fields $\sim n{\rm G}$ and the axion mass $m_a >10^{-22}{\rm eV}$,
we can neglect the effect of magnetic fields $\beta \ll 1$.
However, for more strong magnetic fields,
we need to consider the effect of $\beta$.

In the end, there are uniform static magnetic field, oscillating axion field and oscillating electric field in the background:
\begin{align}
	&a_0(\tau)=\bar{a}\cos\left(\Omega_\beta\,\tau\right)\ ,\\
	&E_{0\parallel}(\tau)=m_a\beta\,\bar{a}\cos\left(\Omega_\beta\,\tau\right)\ .\
\end{align}
We introduce energy density $\rho$ as follows:
\begin{align}
	\rho\equiv\left.\frac{1}{2}(\partial_t\, a_0)^2 + \frac{1}{2}m^2_a\,a^2_0\right|_{\rm present}
	+\frac{1}{2}\,E^2_{0\parallel}(\tau)=\frac{1}{2}\,\bar{a}^2m^2_a\,\Omega^2_\beta\ ,
\end{align}
then background energy density $\rho_{\rm BG}$ is given by
\begin{align}
	\rho_{\rm BG} \equiv \rho+\frac{1}{2}\,B^2_0\ .
\end{align}
We determine axion amplitude $\bar{a}$ by the energy density  $\rho$,
\begin{align}
	\bar{a} = \frac{\sqrt{2\rho}}{m_a\Omega_\beta}\ .
\end{align}
Thus, we found the following expressions
\begin{align}
	&\label{eq:background_axion}
	a_0(\tau)=\frac{\sqrt{2\rho}}{m_a\,\Omega_\beta}\cos\left(\Omega_\beta\,\tau\right)\ ,\\
	&\label{eq:background_electric_field}
	E_{0\parallel}(\tau)=
	\frac{\beta\sqrt{2\rho}}{\Omega_\beta}
	\cos\left(\Omega_\beta\,\tau\right)\ .
\end{align}

\subsection{\label{subsec:Perturbative_Equations}Perturbative Equations}
Now, let us divide the fields into background and perturbation as follows:
\begin{align}\label{eq:zeromode_nonzeromode}
	\begin{cases}
		&a(z, t) = \ a_0 (t) \ \ \ \ \ \ \ \ \ + \delta a(z, t)\ ,\\
		&\vec{B} (z, t) = \ \vec{B}_0\,({\rm const.}) \ + \delta \vec{B} (z, t)\ ,\\
		&\vec{E}(z, t) = \ \vec{E}_0(t) \ \ \ \ \ \ \ \ + \delta \vec{E} (z, t)\ .
	\end{cases}
\end{align}
The first order equations of (\ref{eq:Klein-gordon}) and (\ref{eq:Maxwell}) are given by
\begin{align}
	&(\Box-m^2_a)\,\delta a(z, t)=g_{a\gamma\gamma}\,
	\left[\,\delta\vec{E}(z, t)\cdot\vec{B}_0 
	+ \vec{E}_0(t)\cdot\delta\vec{B}(z, t)\,\right]\ ,
\end{align}
and 
\begin{align}
	\begin{cases}
	\Box\,\delta\phi(z, t)=-g_{a\gamma\gamma}\,\vec{B}_0\cdot[\nabla \delta a(z, t)\,]\ ,\\
	\Box\,\delta\vec{A}(z, t)=g_{a\gamma\gamma}\,
	\left[\left[\partial_t\,\delta{a}(z, t)\right]\vec{B}_0
	+\left[\partial_t\,{a}_0(t)\right]\delta\vec{B}(z, t)+\left[\nabla\,\delta a(z, t)\right]\times\vec{E}_0(t)\,\right]\ .
	\end{cases}
\end{align}
In our set up,
the background magnetic field has only $\parallel$-component,
and the propagating direction of axion is $z$-axis.
Hence, the source term of the scalar potential $\delta \phi(z, t)$ equation vanishes
\begin{align}
	\Box\,\delta \phi(z, t) = 0  \ .
\end{align}
Thus, we can choose the radiation gauge
\begin{align}
	 \delta\phi(z, t) = 0,\ \ \nabla \delta\vec{A}(z, t) =0\ .
\end{align}
In terms of components,
we can write the equations as follows:
\begin{align}
	&(\Box-m^2_a)\delta a(z, t)=g_{a\gamma\gamma}\left[\delta E_\parallel(z, t)B_0-E_{0\parallel}(t)\frac{\partial\,\delta A_{\perp}(z, t)}{\partial\,z}\right]\ ,\\	
	&\Box\,\delta A_\parallel(z, t)=g_{a\gamma\gamma}
	\left[\left[\partial_t\,\delta a(z, t)\right]B_0-\left[\partial_t\,a_{0}(t)\right]\frac{\partial\,\delta A_\perp(z, t)}{\partial\,z}\right]\ ,\\
	&\Box\,\delta A_\perp(z, t)=g_{a\gamma\gamma}
	\left[\left[\partial_t\,a_{0}(t)\right]\frac{\partial\,\delta A_\parallel(z, t)}{\partial z}
	+\frac{\partial\,\delta a(z, t)}{\partial z}E_{0\parallel}(t)\right]\ ,\\
	&\Box\,\delta A_z(z, t)=-g_{a\gamma\gamma}\frac{\partial\,\delta a(z, t)}{\partial y}E_{0\parallel}(t)\ .
\end{align}

Although the time translational symmetry is broken by the time dependent coherent oscillation of the axion field,
the system has the spatial translation invariance.
Hence, it is useful to use Fourier transformation
\begin{align}
	&\delta a(z, t)=\int \frac{dk}{2\pi}\,\delta a(k, t)\,e^{ikz}\ ,\\
	&\delta A_{\alpha}(z, t) = \int \frac{dk}{2\pi}\,\delta A_{\alpha}(k, t) \,e^{ikz}\ ,\\
	&\delta E_{\alpha}(z, t) = - \int \frac{dk}{2\pi}\,\left[\partial_t\,\delta A_{\alpha}(k, t)\right]\,e^{ikz}\ ,
\end{align}
where $\alpha$ denotes $\parallel$ or $\perp$.
Using this transformation, we can write the equations
as follows:
\begin{align}
	&\partial^2_t\,{\delta a}(k, t) + (k^2 + m^2_a)\,\delta a(k, t)
	=g_{a\gamma\gamma}\,B_0\,\left[\partial_t\,{\delta A}_\parallel(k, t)\right]
	+i\,g_{a\gamma\gamma}\,k\,E_{0\parallel}(t)\,\delta A_\perp(k, t)\label{eq:nonzero_full_axion}\ ,\\
	&\partial^2_t\,{\delta A}_\parallel(k, t)+k^2\,\delta A_\parallel(k, t)
	=-g_{a\gamma\gamma}\,B_0\,\left[\partial_t\,{\delta a}(k, t)\right]
	+i\,g_{a\gamma\gamma}\,k\left[\partial_t\,{a}_0(t)\right]\,\delta A_{\perp}(k, t)\label{eq:nonzero_full_parallel}\ ,\\
	&\partial^2_t\,{\delta A}_\perp(k, t)+k^2\,\delta A_\perp(k, t)
	=-i\,g_{a\gamma\gamma}\,k\,\left[\partial_t\,{a}_0(t)\right]\,\delta A_\parallel(k, t)
	-i\,g_{a\gamma\gamma}\,k\,E_{0\parallel}(t)\,\delta a(k, t)
	\label{eq:nonzero_full_perp}\ .
\end{align}
Now, we need to substitute the background solutions (\ref{eq:background_axion}) and (\ref{eq:background_electric_field})
into Eqs.~(\ref{eq:nonzero_full_axion})--(\ref{eq:nonzero_full_perp}).
Then, we get following equations:
\begin{align}
	&\label{eq:full_axion}\delta \ddot{a}(k,\tau)+\left[1+\kappa^2\right]\,\delta a(k,\tau)
	=\,\beta\,\delta \dot{A}_\parallel(k,\tau)+i\frac{\beta\epsilon}{\Omega_\beta}
	\cos\left(\Omega_\beta\tau\right)\delta A_{\perp}(k, t)	\ ,\\
	&\label{eq:full_parallel}\delta \ddot{A}_{\parallel}\,(k,\tau)+\kappa^2\delta A_{\parallel}\,(k,\tau)
	=-\beta\,\delta \dot{a}(k,\tau) 
	-i\epsilon\sin\left(\Omega_\beta\tau\right)\delta A_{\perp}(k, t)\ ,\\
	&\label{eq:full_perp}\delta \ddot{A}_{\perp}\,(k,\tau)+\kappa^2\delta A_{\perp}\,(k,\tau)
	=i\epsilon\sin\left(\Omega_\beta\tau\right)\delta A_{\parallel}\,(k,\tau)-i\frac{\beta\epsilon}{\Omega_\beta}\cos\left(\Omega_\beta\tau\right)\delta a(k,\tau)\ ,
\end{align}
where we introduced dimensionless parameters $\kappa$ and $\epsilon$ as follows:
\begin{align}
	\kappa \equiv \frac{k}{m_a} \ ,  \ \ 
	\epsilon \equiv \frac{g_{a\gamma\gamma}\,\sqrt{2\rho}\,k}{m^2_a}
	=\frac{g_{a\gamma\gamma}\,\sqrt{2\rho}}{m_a}\,\kappa\ .
\end{align}

From now on, for simplicity, we use an approximation neglecting higher order terms
in $\beta$ and $\epsilon$, namely, we take into account 
up to the first order in $\beta$ and $\epsilon$. Thus, we obtain
\begin{align}
	&\delta \ddot{a}(k,\tau)+\left[1+\kappa^2\right]\,\delta a(k,\tau)
	=\beta\,\delta \dot{A}_\parallel(k,\tau)\label{eq:fundamental_axion}\ ,\\
	&\delta \ddot{A}_{\parallel}\,(k,\tau)+\kappa^2\delta A_{\parallel}\,(k,\tau)
	=-\beta\,\delta \dot{a}(k,\tau)-i\,\epsilon\,\sin(\tau)\,\delta A_{\perp}\,(k,\tau)
	\label{eq:fundamental_parallel}\ ,\\
	&\delta \ddot{A}_{\perp}\,(k,\tau)	
	+\kappa^2\,\delta A_{\perp}\,(k,\tau)=i\,\epsilon\,\sin(\tau)\,\delta A_{\parallel}\,(k,\tau)\ .
	\label{eq:fundamental_perp}
\end{align}
We can see 
that when $\epsilon=0$, Eqs.~(\ref{eq:fundamental_axion})--(\ref{eq:fundamental_perp})
 describe the axion--photon conversion
 ~\cite{10.1016/0370-2693(86)90869-5,10.1103/PhysRevD.37.1237}.
For  $\beta=0$, they describe
photon propagation in the presence of only axion dark matter~\cite{10.1093/ptep/pty029}.

Taking the circular polarization basis
\begin{align}
	\vec{e}_{\,L/R}=\frac{1}{\sqrt{2}}\left[\vec{e}_\parallel\mp\,i\,\vec{e}_\perp\right]\ ,
\end{align}
we see original Eqs.~(\ref{eq:full_axion})--(\ref{eq:full_perp})
are rewritten as follows:
\begin{align}
	\delta \ddot{a}(k,\tau)+\left[1+\kappa^2\right]\,\delta a(k,\tau)
	=&\frac{\beta}{\sqrt{2}}\,\left[\delta \dot{A}_L(k,\tau)
	+\delta \dot{A}_R(k,\tau)\right]\nonumber\\
	&-\frac{\beta\epsilon}{\sqrt{2}\,\Omega_\beta}
	\cos\left(\Omega_\beta\,\tau\right)
	\left[-\delta \dot{A}_L(k,\tau)
	+\delta \dot{A}_R(k,\tau)\right]\ ,\label{eq:full_LR_axion}
\end{align}
\begin{align}
	&\delta \ddot{A}_{L/R}\,(k,\tau)
	+\left[\,\kappa^2\pm\,\epsilon\,\sin(\tau)\right]\delta A_{L/R}\,(k,\tau)
	=-\frac{\beta}{\sqrt{2}}\,\delta \dot{a}(k,\tau)
	\pm\frac{\beta\epsilon}{\sqrt{2}\,\Omega_\beta}
	\cos\left(\Omega_\beta\,\tau\right)\delta a(k,\tau)\ ,\label{eq:full_LR_photon}
\end{align}
Under the approximation we are considering, Eqs.~(\ref{eq:fundamental_axion})--(\ref{eq:fundamental_perp})
are rewritten as follows:
\begin{align}
	&\delta \ddot{a}(k,\tau)+\left[1+\kappa^2\right]\,\delta a(k,\tau)
	=\frac{\beta}{\sqrt{2}}\,\left[\delta \dot{A}_L(k,\tau)
	+\delta \dot{A}_R(k,\tau)\right]\ ,\label{eq:fundamental_LR_axion}\\
	&\delta \ddot{A}_{L/R}\,(k,\tau)
	+\left[\,\kappa^2\pm\,\epsilon\,\sin(\tau)\right]\delta A_{L/R}\,(k,\tau)
	=-\frac{\beta}{\sqrt{2}}\,\delta \dot{a}(k,\tau)\ .\label{eq:fundamental_LR_photon}
\end{align}

Here, we should mention the previous work~\cite{10.1142/S0217751X15500992}.
They investigated the similar system, but they neglected the parametric resonance.
In this paper, we consider Mathieu type terms and focus on the resonance instability.

\section{\label{sec:Ince--Strutt Chart}Stability Analysis --- Ince--Strutt Chart}
In this section, we numerically investigate the behavior of solutions for the basic equations
Eqs.~(\ref{eq:fundamental_axion})--(\ref{eq:fundamental_perp}).
First, we give a short review of the Mathieu equation for comparison.
Next, we show numerical results for axion dark matter--photon conversion
which have both similarities and differences with the Mathieu equation's.
In the next Sec.~\ref{sec:transition_curves}, we will provide analytical derivation of the numerical results.

\subsection{\label{subsec:Ince--Strutt_Mathieu}Without Background Magnetic Field}
\begin{figure}[H]
\begin{center}
	\includegraphics[width=10cm]{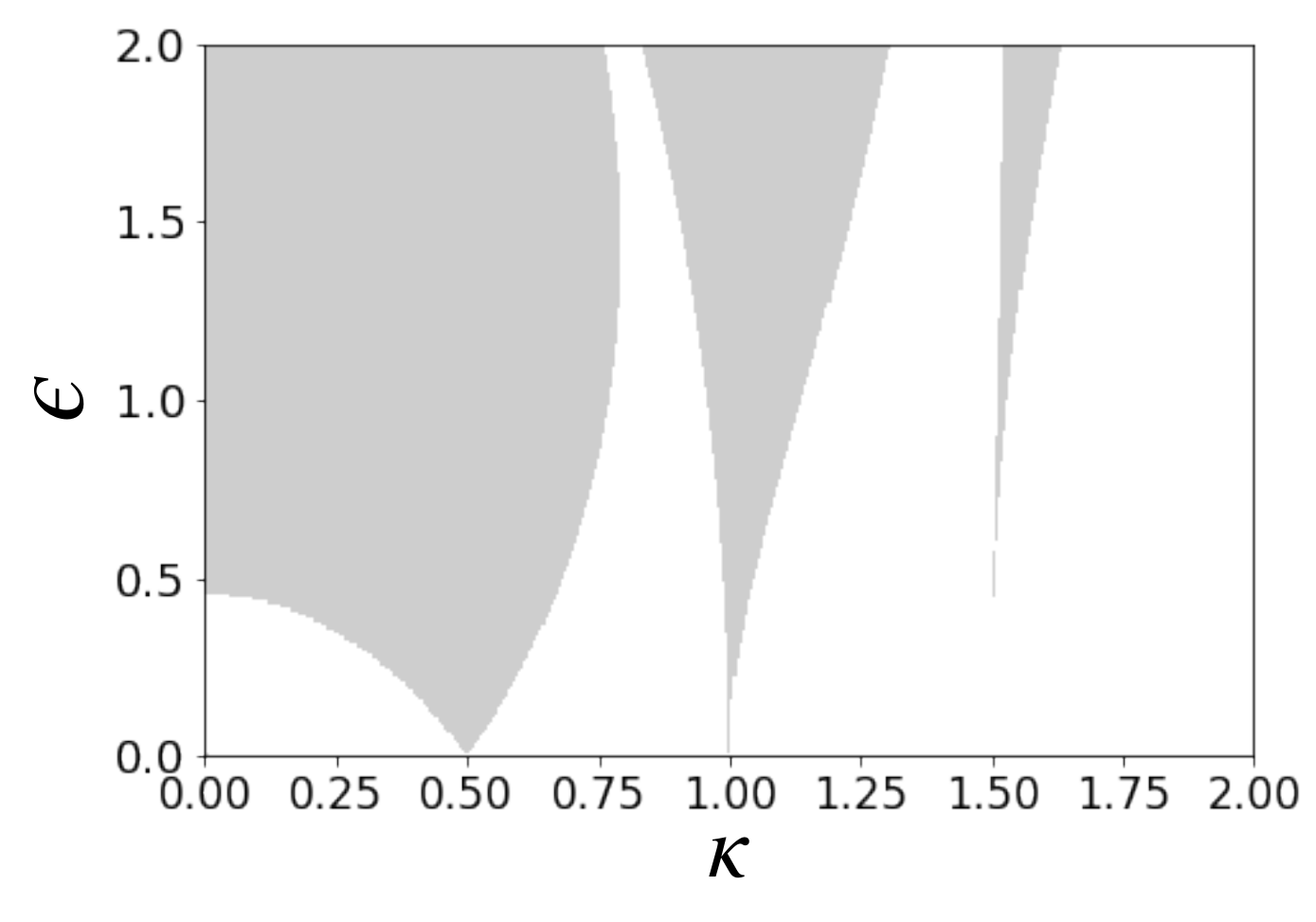}
	\caption{\label{fig:Ince--Strutt_Mathieu}The Ince--Strutt chart for the Mathieu equation $(\bar{\kappa}=n/2)$.
	The shaded area represents parameter set which make solution of the equation (\ref{eq:Mathieu_axionDM}) unstable.}
\end{center}
\end{figure}

A photon propagating in the axion dark matter obeys following equations~\cite{10.1093/ptep/pty029}.
\begin{align}
	\delta \ddot{A}_{L/R}\,(k,\tau)
	+\left[\,\kappa^2\pm\,\epsilon\,\sin(\tau)\right]\delta A_{L/R}\,(k,\tau)=0\ .
	\label{eq:Mathieu_axionDM}
\end{align}
This can be obtained by putting $\beta =0$ in Eq.~(\ref{eq:fundamental_LR_photon}).
The equation (\ref{eq:Mathieu_axionDM}) represents harmonic oscillator
whose frequency also oscillates, and this type of equation
is called the Mathieu equation~\cite{Mathieu1868,978-0486612331,10.1115/1.4039144}.
The solutions can be stable or unstable,
depending on dimensionless parameters, $\kappa$ and $\epsilon$.
The Floquet theorem~\cite{Floquet1883} divide the $(\kappa-\epsilon)$ plane 
into two regions (Fig.~\ref{fig:Ince--Strutt_Mathieu}),
stable and unstable,
and this chart is called Ince--Strutt chart~\cite{10.1017/S0370164600021866,10.1002/andp.19283911006}.
Please refer the reader to  \cite{10.1115/1.4039144} for the Floquet theorem and the Ince--Strutt chart.

The bifurcation points on the $\kappa$ axis appear at
\begin{align}
	\bar{\kappa}\equiv\frac{n}{2}\ \ (n=1, 2, 3,\cdots)\ ,\label{eq:Mathieu_condition}
\end{align}
and the boundaries between the stable and unstable region are called transition curves.
On the transition curves, the Eq.~(\ref{eq:Mathieu_axionDM}) has periodic solutions.
Here, we introduce dimension less parameter $\chi$:
\begin{align}
	\kappa =\bar{\kappa} + \chi\ ,\label{def:tildechi_chi}
\end{align}
where $\bar{\kappa}$ is given by Eq.~(\ref{eq:Mathieu_condition}).
For nonzero $\epsilon\neq0$,
a wave number $\kappa$ deviates from $\bar{\kappa}$
in order for the solution of Eq.~(\ref{eq:Mathieu_axionDM})
to still have a period $T= 2\pi/\bar{\kappa}$,
and the deviation is represented by $\chi$.

For example, on the transition curves originated at $\bar{\kappa}=1/2$, 
there is a periodic solution with $T=4\pi$.
For small $|\epsilon|$, the transition curves are approximately given by
\begin{align}
	\chi=\pm\frac{\epsilon}{2}\ .
	\label{eq:Mathieu_transition_05}
\end{align}
In the case of $\bar{\kappa}=1$,
the transition curves are given by
\begin{align}
	\chi_{-}\equiv-\frac{\epsilon^2}{24}\ ,\ \ \chi_{+}\equiv\frac{5\epsilon^2}{24}\ .
	\label{eq:Mathieu_transition_1}
\end{align}
On these curves, (\ref{eq:Mathieu_axionDM}) has a periodic solution with $T=2\pi$.

In the following two sections, 
we will investigate how these results are changed 
when the background magnetic field is taken into account
by solving Eqs.~(\ref{eq:fundamental_axion})--(\ref{eq:fundamental_perp}) numerically.
At the same time, we show some transition curves, 
and its analytical derivation is given in Sec.~\ref{sec:transition_curves}.

\subsection{\label{subsec:shifted}Shift of Bifurcation Points}
\begin{figure}[ht]
\begin{center}
\begin{minipage}[b]{0.46\linewidth}
\includegraphics[keepaspectratio, scale=0.46]{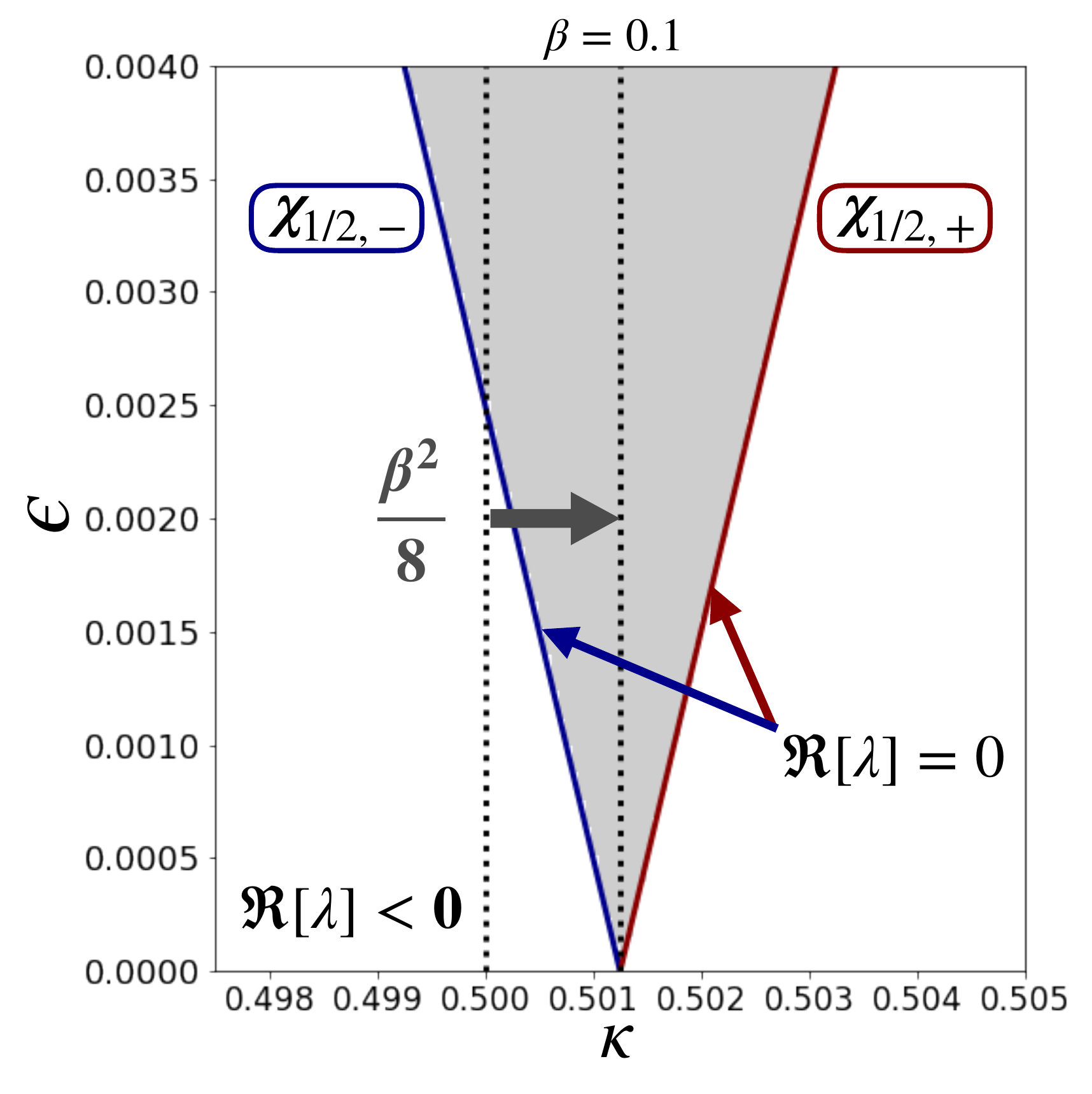}
\subcaption{\label{fig:shifted_05}$\bar{\kappa}=1/2$}
\end{minipage}
\begin{minipage}[b]{0.44\linewidth}
\includegraphics[keepaspectratio, scale=0.5]{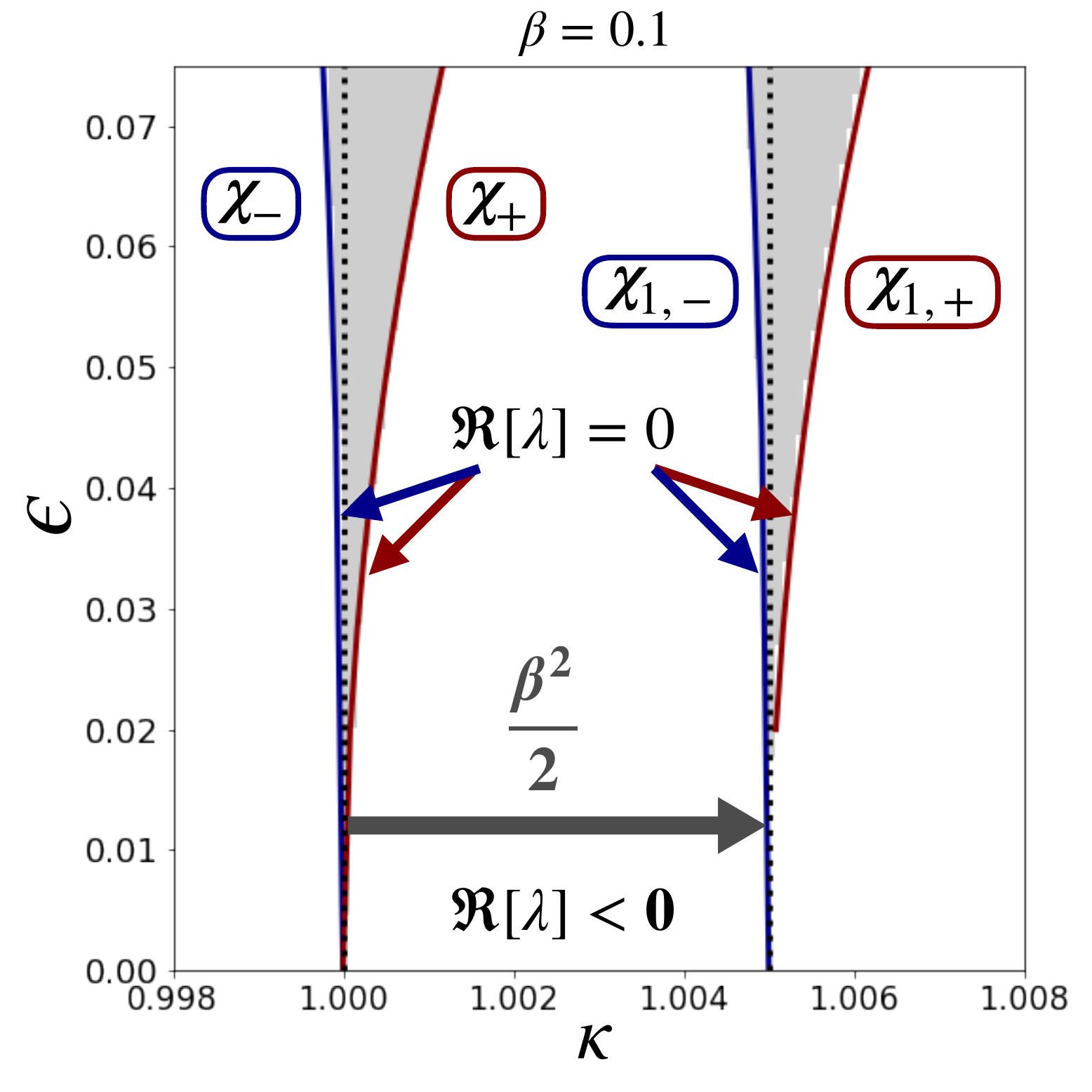}
\subcaption{\label{fig:shifted_1}$\bar{\kappa}=1$}
\end{minipage}
\end{center}
\caption{\label{fig:Ince--Strutt_Shifted}The Ince--Strutt chart for axion DM--photon conversion. 
The shaded area represents parameter set which make solution of Eqs.~(\ref{eq:fundamental_axion})--(\ref{eq:fundamental_perp}) unstable.}
\end{figure}

The Fig.~\ref{fig:Ince--Strutt_Shifted} shows that 
bifurcation points of transition curves appear again around $\bar{\kappa}=n/2\ (n=1, 2, 3,\cdots)$.
To be more precise, bifurcation points are shifted even on the $\kappa$ axis ($\epsilon=0$)
due to the background magnetic field.

As can be seen from Fig.~\ref{fig:shifted_05}, 
the starting point of transition curves $\bar{\kappa}\sim1/2$ is shifted by magnetic field
as
\begin{align}
	&\chi_{1/2,-}\equiv-\frac{\epsilon}{2}+\frac{\beta^2}{8}\label{eq:chi_{1/2,-}}\ ,\\
	&\chi_{1/2,+}\equiv\frac{\epsilon}{2}+\frac{\beta^2}{8}\label{eq:chi_{1/2,+}}\ .
\end{align}
In the case of $\bar{\kappa}\sim1$ (Fig.~\ref{fig:shifted_1}),
the unstable region splits into two regions.
\begin{align}
	&\chi_{-}\equiv-\frac{\epsilon^2}{24}\label{eq:chi_{-}}\ ,\\
	&\chi_{+}\equiv\frac{5\epsilon^2}{24}\label{eq:chi_{+}}\ ,\\
	&\chi_{1,-}\equiv-\frac{\epsilon^2}{24}+\frac{\beta^2}{2}\label{eq:chi_{1,-}}\ ,\\
	&\chi_{1,+}\equiv\frac{5\epsilon^2}{24}+\frac{\beta^2}{2}\label{eq:chi_{1,+}}\ .
\end{align}
The first two curves are exactly the same as the conventional one (\ref{eq:Mathieu_transition_1}).
On the other hand, the other two curves are shifted by magnetic field $\beta$.
The region which intervene between $\chi_{1,-}$ and $\chi_{1,+}$
represents the instability of parallel photon component 
which does interact with axion through magnetic field.

We verified that transition curves (\ref{eq:chi_{1/2,-}})--(\ref{eq:chi_{1/2,+}}) and (\ref{eq:chi_{-}})--(\ref{eq:chi_{1,+}})
are still valid for full equations~(\ref{eq:full_axion})--(\ref{eq:full_perp}).
However, in the case of (\ref{eq:full_axion})--(\ref{eq:full_perp}),
a new bifurcation point appears between $\kappa=1$ and $\kappa =1+\beta^2/2$
due to higher order contributions.

\subsection{\label{subsec:new}New Bifurcation Points}
\begin{figure}[H]
\begin{center}
	\includegraphics[width=9cm]{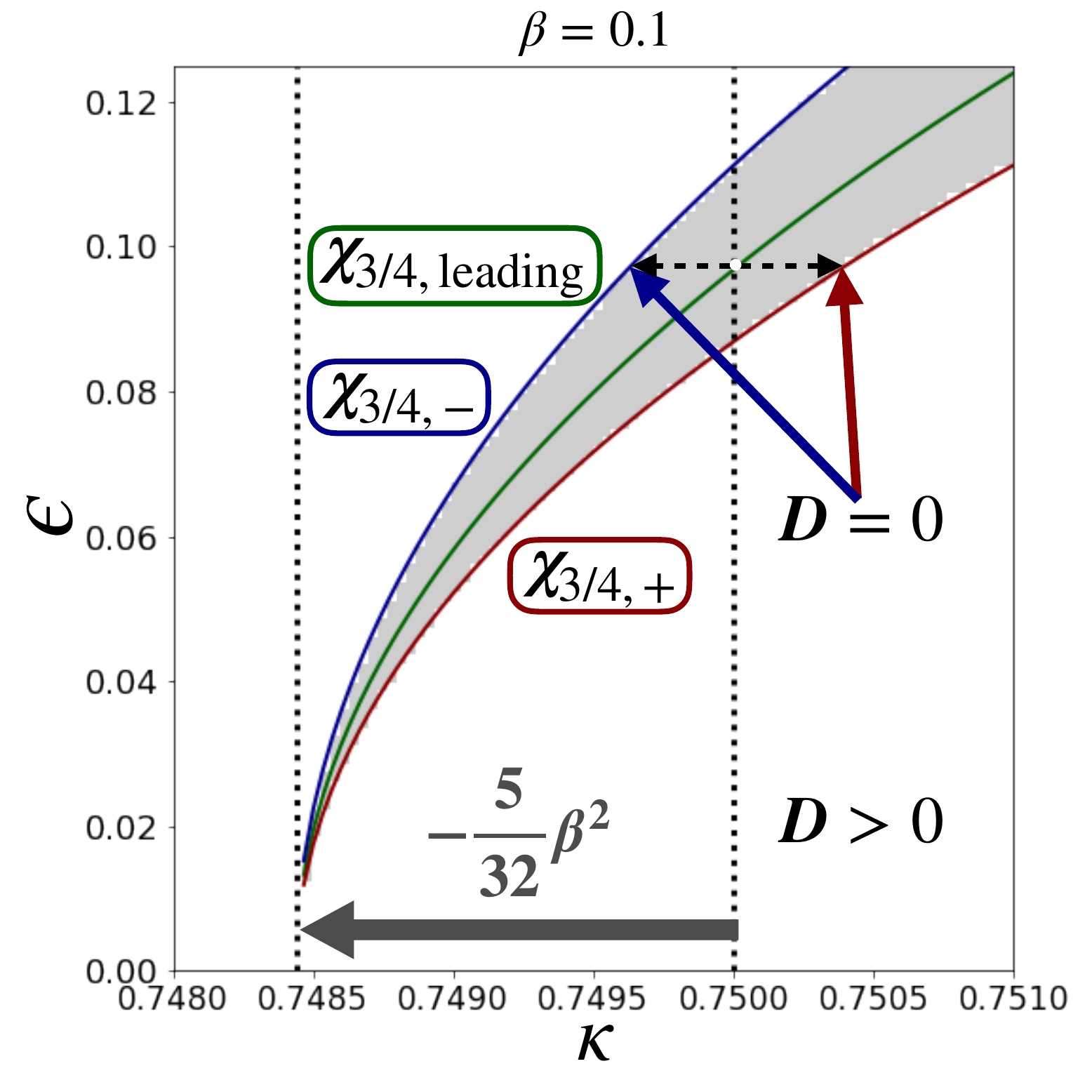}
	\caption{\label{fig:new_075}Axion DM--photon conversion in the case of $\bar{\kappa}=3/4$.
The shaded area represents parameter set which make solution of the equation
 (\ref{eq:fundamental_axion})--(\ref{eq:fundamental_perp}) unstable.
 $D$ in this figure means $D_{\rm up\ to\ next\ leading}$ in Sec.~\ref{subsec:kappa_075}.}
\end{center}
\end{figure}

It seems that the axion dark matter--photon conversion has other bifurcation points.
From our numerical calculations, 
we empirically found the condition for the bifurcation points
\begin{align}
	\sqrt{1+\bar{\kappa}^2} + \bar{\kappa}=n\ \ (n=2, 3, 4, \cdots)\ .
\end{align}
Solving this with respect to $\bar{\kappa}$,
we get the following relation:
\begin{align}
	\bar{\kappa} \equiv\frac{n^2-1}{2n}\ \ (n=2, 3, 4, \cdots)\ .
\end{align}
In the case of $n=2$, we depicted the unstable region in Fig.~\ref{fig:new_075}. 
We will see that transition curves can be  derived in an analytical way in the next section
as follows:
\begin{align}
	&\chi_{3/4,{\rm leading}}\equiv-\frac{5}{32}\,\beta^2+\frac{\epsilon^2}{6}
	\label{eq:chi_075_leading}\ ,\\
	&\chi_{3/4, -}\equiv\chi_{3/4,{\rm leading}}-\frac{5}{48}\sqrt{15}\beta\epsilon^2
	\label{eq:chi_075_leading_-}\ ,\\
	&\chi_{3/4, +}\equiv\chi_{3/4,{\rm leading}}+\frac{5}{48}\sqrt{15}\beta\epsilon^2
	\label{eq:chi_075_leading_+}\ .
\end{align}

We checked that the Eq.~(\ref{eq:chi_075_leading}) is still valid for full equations~(\ref{eq:full_axion})--(\ref{eq:full_perp}).
However, in the case of (\ref{eq:full_axion})--(\ref{eq:full_perp}),
the width of unstable band is more broader due to more higher order contributions.
Moreover, a new bifurcation point appears on the left side of $\kappa=3/4-5\beta^2/32$.
In this paper, we shall restrict ourselves 
to the instability of the leading order equations (\ref{eq:fundamental_axion})--(\ref{eq:fundamental_perp}).

\section{\label{sec:transition_curves}Analytic Expressions of Transition Curves}
As we have seen in the previous section, there are differences 
between the conventional Mathieu equation and axion dark matter--photon conversion.
However, the results are obtained numerically.
In this section, we would like to give an analytical support to our findings.
We show how the boundaries between stable and unstable regions are determined
by treating parameters $\chi,\,\beta,\,\epsilon$ as small quantities.

The basic equations 
(\ref{eq:fundamental_axion})--(\ref{eq:fundamental_perp})
can be written as follows:
\begin{align}\label{eq:fundamental_matrix}
		\ddot{\vec{x}} + B\,\dot{\vec{x}} + [K + E \sin(\tau)]\,{\vec{x}} = 0\ ,
\end{align}
by using vector and matrices, 
\begin{align}\label{eq:vec_matrix}
	\vec{x}=
	\begin{bmatrix}
		\delta a (k,\tau)\\
		\delta A_\parallel (k,\tau)\\
		\delta A_\perp (k,\tau)
	\end{bmatrix},\ \ \ 
	B = 
	\begin{bmatrix}
		0 & -\beta & 0\\ 
		\beta & 0  & 0\\
		0 &  0 & 0
	\end{bmatrix},\ \ \ 
	K=
	\begin{bmatrix}
		1+\kappa^2 & 0 &0\\
		0 & \kappa^2&0 \\
		0 & 0 & \kappa^2
	\end{bmatrix},\ \ \ 
	E =
	\begin{bmatrix}
		0 & 0 & 0\\
		0 & 0 & i \epsilon\\
		0 & -i\epsilon & 0
	\end{bmatrix}\ .
\end{align}
Let us put an ansatz 
\begin{align*}
	\vec{x}(\tau) = e^{\lambda\tau}
	\left[\ 
	\text{The superpositions of many photon's overtones.}\ 
	\right]\ ,
\end{align*}
and substitute it into to the Eq.~(\ref{eq:fundamental_matrix}).
In order for (\ref{eq:fundamental_matrix}) to have a nontrivial solution, {\it i.e.}, $\vec{x}\neq0$,
the determinant of the coefficient matrix obtained in this way must vanish.
Here, we introduce growth rate $\lambda\in\mathbb{C}$
under the condition, $|\chi|\sim|\lambda|\ll1$.
It is the real part $\Re[\lambda]$ that determines the stability of the solutions to the Eq.~(\ref{eq:fundamental_matrix}) ,
and the imaginary part $\Im[\lambda]$ detune the frequency of the solutions.
The criteria of the stable and unstable is given as follows:
\begin{align}\label{eq:condition_lambda}
	\Re[\lambda]\leq0\ \text{: stable},\ \ \Re[\lambda]>0\ \text{: unstable}.
\end{align}
In the case of $\Re[\lambda]>0$, 
the growth rate after one period $T$ is given by roughly $e^{T\Re[\lambda]}$,
\begin{align}
	\vec{x}(\tau+T)\simeq e^{T\Re[\lambda]}\,\vec{x}(\tau)\ .
\end{align}
From the explicit calculations,
it turns out that the determinant of the coefficient matrix depends only on 
$q\equiv\lambda^2$. 
Therefore, the criteria (\ref{eq:condition_lambda}) can be replaced by following ones:
\begin{align}\label{eq:condition_q}
	\Im[q\equiv\lambda^2]=0\ \text{and}\ \Re[q\equiv\lambda^2]\leq0\ \text{: stable}
	,\ \ \text{all other condition: unstable}.
\end{align}
Note that small parameters $\chi,\,\beta,\,\,\epsilon,\,\lambda$
may have different relative magnitude relationship
depending on  bifurcation points.

Before moving on to the concrete analysis, we define the $3\times3$ matrices
which compose coefficient matrix:
\begin{align}\label{}
	&\text{Diag}(n)\equiv \lambda^2I_3+\lambda B+K-(\kappa n)^2\,I_3\ ,\label{eq:diag_05_1}\\
	&\text{Mix}(n)\equiv4\kappa n\left[\frac{\lambda}{2}\,I_3+\frac{B}{4}\right]\ ,\label{eq:mix_05_1}
\end{align}
where $I_3$ denotes the identity matrix, 
and $n$ is non-negative integer, $n=0, 1, 2, \cdots$.
The matrices $K, B, E$ have  been already defined in (\ref{eq:vec_matrix}).

\subsection{\label{subsec:kappa_05} Shift of Bifurcation Point at $\bar{\kappa} = 1/2$}
Substituting the ansatz
\begin{align}
	\vec{x}(\tau) = e^{\lambda\tau}
	\left[\,
	\vec{a}_1\cos\left(\frac{\tau}{2}\right)+\vec{b}_1\sin\left(\frac{\tau}{2}\right)
	\,\right]\ ,
\end{align}
into the Eq.~(\ref{eq:fundamental_matrix}),
we obtain $6\times6$ coefficient matrix $R_{1/2}(q, \chi, \beta, \epsilon)$.
\begin{align}
	R_{1/2}(q, \chi, \beta, \epsilon)\equiv
	\begin{bmatrix}
		\text{Diag}(1)&\text{Mix}(1)+E/2\\
		-\text{Mix}(1)+E/2&\text{Diag}(1)\\
	\end{bmatrix}\ .
\end{align}
The determinant of $R_{1/2}(q, \chi, \beta, \epsilon)$ must vanish.
\begin{align}
	{\rm det}[R_{1/2}(q, \chi, \beta, \epsilon)]=0\ .
\end{align}
From numerical results, we see the hierarchy of the order  $\lambda\sim\chi\sim \epsilon \sim \beta^2$. 
The leading order contribution to the determinant is given by 
\begin{align}
	\label{eq:05_det_concleate_leading}
	&{\rm det}[R_{1/2}(q, \chi, \beta, \epsilon)]_{\rm leading}\nonumber\\
	&=\left[q-\left(\sqrt{-\left(\chi-\frac{\beta^2}{8}\right)^2+\frac{\epsilon^2}{4}}+i\frac{\beta^2}{8}\right)^2\right]
	\left[q-\left(\sqrt{-\left(\chi-\frac{\beta^2}{8}\right)^2+\frac{\epsilon^2}{4}}-i\frac{\beta^2}{8}\right)^2\right]\ .
\end{align}
The next leading order is not relevant here, however,
you will soon see that it should be taken into account 
when you consider the new bifurcation points. 
In the present case, we have the following next order contribution:
\begin{align}
	\label{eq:05_det_concleate_next_leading}
	{\rm det}[R_{1/2}(q, \chi, \beta, \epsilon)]_{\rm next\ leading}
	&=(2\beta^2+6\chi)q^2
	+\left(12\chi^3-\frac{1}{2}\epsilon^2\beta^2-2\chi\epsilon^2+\frac{1}{8}\beta^4\chi\right)q \nonumber \\
	&\quad +\frac{1}{8}\epsilon^4\chi-2\beta^2\chi^4-2\chi^3\epsilon^2+\frac{1}{8}\beta^4\chi^3+6\chi^5+\frac{1}{4}\epsilon^2\beta^2\chi^2\ .
\end{align}
Evaluating the determinant at the leading order
\begin{align}
	{\rm det}[R_{1/2}(q, \chi, \beta, \epsilon)]_{\rm leading}=0\ ,
\end{align}
 we get four $\lambda$:
\begin{align}\label{eq:growth_rate_05}
	\lambda=\pm\sqrt{-\left(\chi-\frac{\beta^2}{8}\right)^2+\frac{\epsilon^2}{4}}
	\pm i\frac{\beta^2}{8}\ ,\ \ 
	\pm\sqrt{-\left(\chi-\frac{\beta^2}{8}\right)^2+\frac{\epsilon^2}{4}}
	\mp i\frac{\beta^2}{8}\ .
\end{align}
Therefore, the range where the criterion for stability (\ref{eq:condition_lambda}) is broken is as follows:
\begin{align}
	-\frac{\epsilon}{2}<\chi-\frac{\beta^2}{8}<\frac{\epsilon}{2}\ ,
\end{align}
and the transition curves are derived from the condition $\Re[\lambda]=0$\ ,
\begin{align}
	\label{eq:05_transition}
	\chi-\frac{\beta^2}{8}=\pm \frac{\epsilon}{2}\ .
\end{align}
Here, we would like to comment on the growth rate (\ref{eq:growth_rate_05}).
If you consider the situation where there is only coherent oscillating axion in the background like~\cite{10.1093/ptep/pty029,10.1088/1475-7516/2018/11/004,10.1140/epjc/s10052-019-6759-7},
then the growth rate does not depend on axion mass $m_a$.
In fact, if you choose $\beta=0$ in (\ref{eq:growth_rate_05}), then you can confirm this fact.
However, in our case $\beta\neq0$, note that growth rate $m_a\Re[\lambda]$ become to depend on axion mass  due to the presence of background magnetic field $\beta$.

\subsection{\label{subsec:kappa_1}Shift of Bifurcation Point at $\bar{\kappa} = 1$}
Substituting the ansatz
\begin{align}
	\vec{x}(\tau) = e^{\lambda\tau}
	\left[\,
	\vec{a}_1\cos\left(\tau\right)+\vec{b}_1\sin\left(\tau\right)
	+\vec{a}_2\cos\left(2\tau\right)+\vec{b}_2\sin\left(2\tau\right)
	+\vec{c}
	\,\right]\ ,
\end{align}
into the Eq.~(\ref{eq:fundamental_matrix}),
we obtain $15\times15$ coefficient matrix $R_1(q, \chi, \beta, \epsilon)$.
\begin{align}
	R_1(q, \chi, \beta, \epsilon)\equiv
	\begin{bmatrix}
		\text{Diag}(0)&0&E/2&0&0\\
		0&\text{Diag}(1)&\text{Mix}(1)&0&E/2\\
		E&-\text{Mix}(1)&\text{Diag}(1)&-E/2&0\\
		0&0&-E/2&\text{Diag}(2)&\text{Mix}(2)\\
		0&E/2&0&-\text{Mix}(2)&\text{Diag}(2)
	\end{bmatrix}\ .
\end{align}
The determinant of $R_1(q, \chi, \beta, \epsilon)$ must vanish.
\begin{align}
	{\rm det}[R_{1}(q, \chi, \beta, \epsilon)]=0\ .
\end{align}
From numerical results, we see the hierarchy of the order $\lambda\sim\chi\sim \epsilon^2 \sim \beta^2$. 
Evaluating the determinant at the leading order
\begin{align}
	{\rm det}[R_1(q, \chi, \beta, \epsilon)]_{\rm leading}=0\ ,
\end{align}
we get a quadratic equation with respect to $q\equiv\lambda^2$,
\begin{align}\label{eq:1_determinant_leading}
	\left[q+\left(\chi+\frac{\epsilon^2}{24}\right)\left(\chi-\frac{5\epsilon^2}{24}\right)\right]
	\left[q+\left(\chi-\frac{\beta^2}{2}+\frac{\epsilon^2}{24}\right)\left(\chi-\dfrac{\beta^2}{2}-\dfrac{5\epsilon^2}{24}\right)\right]=0\ .
\end{align}
Therefore, the range where the criterion for stability (\ref{eq:condition_q}) is broken is as follows:
\begin{align}
	-\frac{\epsilon^2}{24}<\chi<\frac{5\epsilon^2}{24}\ ,
	\ -\frac{\epsilon^2}{24}<\chi-\frac{\beta^2}{2}<\frac{5\epsilon^2}{24}\ ,
\end{align}
and the transition curves are derived 
from the condition $q=0$, {\it i.e.} $\Re[\lambda]=0$,
\begin{align}
	\chi=-\dfrac{\epsilon^2}{24},\ \ \dfrac{5\epsilon^2}{24}\ ,
	\ \ \dfrac{\beta^2}{2}-\dfrac{\epsilon^2}{24},\ \ \dfrac{\beta^2}{2}+\dfrac{5\epsilon^2}{24}\ .
\end{align}

\subsection{\label{subsec:kappa_075}A New Bifurcation Point at $\bar{\kappa} = 3/4$}
This is a new unstable region around $\bar{\kappa}=3/4$
where the conventional Mathieu equation does not have the instability.
Substituting the ansatz
\begin{align}\label{eq:ansatz_3/4}
	\vec{x}(\tau) = e^{\lambda\tau}
	\begin{bmatrix}
		&\ \vec{a}_1\cos\left(\tau/4\right)
			+\vec{b}_1\sin\left(\tau/4\right)
			+\vec{a}_2\cos\left(3\tau/4\right)
			+\vec{b}_2\sin\left(3\tau/4\right)\\
		&\ +\, \vec{a}_3\cos\left(5\tau/4\right)
			+\vec{b}_3\sin\left(5\tau/4\right)
			+\vec{a}_4\cos\left(7\tau/4\right)
			+\vec{b}_4\sin\left(7\tau/4\right)\ 
	\end{bmatrix}\ ,
\end{align}
into Eq.~(\ref{eq:fundamental_matrix}),
we obtain $24\times24$ coefficient matrix $R_{3/4}(q, \chi, \beta, \epsilon)$.
\begin{align}
	&R_{3/4}(q, \chi, \beta, \epsilon)\nonumber\\
	&\equiv
	\begin{bmatrix}
		\text{Diag}_{3/4}(0)&\text{Mix}_{3/4}(0)&0&E/2&0&E/2&0&0\\
		-\text{Mix}_{3/4}(0)&\text{Diag}_{3/4}(0)&E/2&0&-E/2&0&0&0\\
		0&E/2&\text{Diag}_{3/4}(1)&\text{Mix}_{3/4}(1)&0&0&0&E/2\\
		E/2&0&-\text{Mix}_{3/4}(1)&\text{Diag}_{3/4}(1)&0&0&-E/2&0\\
		0&-E/2&0&0&\text{Diag}_{3/4}(2)&\text{Mix}_{3/4}(2)&0&0\\
		E/2&0&0&0&-\text{Mix}_{3/4}(2)&\text{Diag}_{3/4}(2)&0&0\\
		0&0&0&-E/2&0&0&\text{Diag}_{3/4}(3)&\text{Mix}_{3/4}(3)\\
		0&0&E/2&0&0&0&-\text{Mix}_{3/4}(3)&\text{Diag}_{3/4}(3)
	\end{bmatrix}\ .
\end{align}
Note that in this case the definition of matrices are different from
(\ref{eq:diag_05_1}) and (\ref{eq:mix_05_1}),
\begin{align}
	&\text{Diag}_{3/4}(n)\equiv \lambda^2I_3+\lambda B+K-\left(\frac{2n+1}{4}\right)^2\,I_3\ ,\label{eq:diag_3/4}\\
	&\text{Mix}_{3/4}(n)\equiv(2n+1)\left[\frac{\lambda}{2}\,I_3+\frac{B}{4}\right]\ .\label{eq:mix_3/4}
\end{align}
The determinant of $R_{3/4}(q, \chi, \beta, \epsilon)$ must vanish.
\begin{align}
	{\rm det}[R_{3/4}(q, \chi, \beta, \epsilon)]=0\ .
\end{align}
From numerical results, we see the hierarchy of the order $\lambda\sim\chi\sim \epsilon^2 \sim \beta^2$. 
Evaluating the determinant at the leading order
\begin{align}\label{eq:3/4_determinant_leading}
	{\rm det}[R_{3/4}(q, \chi, \beta, \epsilon)]_{\rm leading}=0\ ,
\end{align}
we get a cubic equation for $q\equiv\lambda^2$,
\begin{align}
	\left[q+\frac{9}{25}\left(\chi+\frac{25}{24}\beta^2\right)^2\right]
	\left[q+\left(\chi-\frac{3}{8}\beta^2-\frac{4}{15}\epsilon^2\right)^2\right]
	\left[q+\left(\chi-\frac{4}{15}\epsilon^2\right)^2\right]=0\ .\label{eq:cubic_equation_Dpositive}
\end{align}
We obtained solutions as follows:
\begin{align}
	&q_1\equiv-\frac{9}{25}\left(\chi+\frac{25}{24}\beta^2\right)^2\ ,\\
	&q_2\equiv-\left(\chi-\frac{3}{8}\beta^2-\frac{4}{15}\epsilon^2\right)^2\ ,\\
	&q_3\equiv-\left(\chi-\frac{4}{15}\epsilon^2\right)^2\ .
\end{align}
Equation (\ref{eq:cubic_equation_Dpositive}) has three negative real solution $q_1, q_2 , q_3$,
so $\lambda$ must be pure imaginary.
Hence, no instability occurs.
Unlike $\kappa=1/2$ and $1$, all the relations among the parameters
derived from the condition $q=0$,
\begin{align}\label{eq:3/4_condition_q=0}
	& \chi = -\frac{25}{24}\beta^2\ ,\ \ \frac{3}{8}\beta^2+\frac{4}{15}\epsilon^2\ ,
	\ \ \frac{4}{15}\epsilon^2\ ,
\end{align}
do not give transition curves on the $(\kappa-\epsilon)$ plane.

Now, let us go into the cubic equation with respect to $q=\lambda^2$ (\ref{eq:cubic_equation_Dpositive})
in more detail.
We refer to the discriminant of the cubic equation (\ref{eq:cubic_equation_Dpositive}) as $D_{\rm leading}$.
Equation (\ref{eq:cubic_equation_Dpositive}) says $D_{\rm leading}\geq0$ as long as 
the determinant of coefficient matrix is evaluated at leading order.
In the case of $D_{\rm leading}>0$, even if higher order contributions are considered,
it still remain $D_{\rm higher}>0$.
However, if $D_{\rm leading}=0$, the sign of discriminant $D_{\rm higher}$ can be minus due to
higher order contributions.
This means that two $q$ out of three are complex, and that the solution (\ref{eq:ansatz_3/4})
is always unstable.

Before considering higher order contributions, we derive 
the condition that the equation (\ref{eq:cubic_equation_Dpositive})
has multiple roots, {\it i.e.}, $D_{\rm leading}=0$.
There are three possibilities:
\[q_1 =q_2\ ,\ \ \text{or}\ \ q_2=q_3\ ,\ \ \text{or}\ \ q_3=q_1\ .\]
It is expected that instability around $\bar{\kappa}=3/4$
is caused by the coupling of the axion and the photon ($\parallel$)
through the magnetic field, so
$q_1=q_2$ may be meaningful.
Solving the equation
\begin{align}
	-\frac{9}{25}\left(\chi+\frac{25}{24}\beta^2\right)^2
	=
	-\left(\chi-\frac{3}{8}\beta^2-\frac{4}{15}\epsilon^2\right)^2,
\end{align}
we obtain the relation among parameters,
\begin{align}\label{eq:3/4_condition_D=0}
		\chi=-\frac{5}{32}\beta^2+\frac{\epsilon^2}{6}\ .
\end{align}
On the curve that satisfies this relationship (\ref{eq:3/4_condition_D=0}),
the cubic equation (\ref{eq:3/4_determinant_leading}) can be rewritten,
\begin{align}
	\left[\,q+f_1(\beta, \epsilon)\,\right]\left[\,q+f_2(\beta, \epsilon)\,\right]^2=0\ ,
\end{align}
and it has multiple root $q=-f_2(\beta,\epsilon)$,
where $f_1(\beta, \epsilon)$ and $f_2(\beta, \epsilon)$ are positive real functions.
The cubic equation (\ref{eq:3/4_determinant_leading})
does not have complex solutions at the leading order.

In order to clarify the origin of instability, we need to proceed to the next order.
Please refer the reader to the Appendix.~\ref{appendix} for  concrete expressions.
The determinant at next leading order
${\rm det}[R_{3/4}(q, \chi, \beta, \epsilon)]_{\rm next\ leading} $ is also cubic equation.
Up to the next leading order, we have
\begin{align}
	\label{eq:075_det_concleate_up_to_next_leading}
	{\rm det}&[R_{3/4}(q, \chi, \beta, \epsilon)]_{\rm up\ to\ next \ leading}\nonumber\\
	&\equiv{\rm det}[R_{3/4}(q, \chi, \beta, \epsilon)]_{\rm leading}
	+{\rm det}[R_{3/4}(q, \chi, \beta, \epsilon)]_{\rm next\ leading}    \nonumber\\
	&\equiv C_3(\chi, \beta, \epsilon)q^3+C_2(\chi, \beta, \epsilon)q^2
	+C_1(\chi, \beta, \epsilon)q+C_0(\chi, \beta, \epsilon) = 0 \ ,
\end{align}
where we labeled the coefficients of each orders of $q$ as $C_3 ,C_2 , C_1 ,  C_0$.
The discriminant of cubic equation (\ref{eq:075_det_concleate_up_to_next_leading})
is given by
\begin{align}
	\label{eq:discriminant_def}
	D_{\rm up\ to\ next\ leading}(\chi,\beta,\epsilon)
	\equiv&
	-4C_3(\chi,\beta,\epsilon)[C_1(\chi,\beta,\epsilon)]^3-27[C_3(\chi,\beta,\epsilon)]^2
	[C_0(\chi,\beta,\epsilon)]^2\nonumber\\
	&+[C_2(\chi,\beta,\epsilon)]^2[C_1(\chi,\beta,\epsilon)]^2 -4[C_2(\chi,\beta,\epsilon)]^3C_0(\chi,\beta,\epsilon)  \nonumber\\
	&+18C_3(\chi,\beta,\epsilon)C_2(\chi,\beta,\epsilon)C_1(\chi,\beta,\epsilon)C_0(\chi,\beta,\epsilon)\ .
\end{align}
Let us find a correction term $X$  to the relation among parameters,
\begin{align}\label{eq:3/4_third_order_correction}
	\chi=-\frac{5}{32}\beta^2+\frac{\epsilon^2}{6}+ X\ .
\end{align}
On the curve that satisfies this relationship (\ref{eq:3/4_third_order_correction}),
we expect that the cubic equation (\ref{eq:075_det_concleate_up_to_next_leading}) will have multiple roots,
\begin{align}\label{eq:3/4_determinant_next_leading_D=0}
	\left[\,q+g_1(\beta, \epsilon)\,\right]\left[\,q+g_2(\beta, \epsilon)\,\right]^2=0\ .
\end{align}
In the Eq.~(\ref{eq:3/4_third_order_correction}),
a higher order contribution $X$ is incorporated 
into the leading order relation (\ref{eq:3/4_condition_D=0}).
The correction $X$ is chosen so that 
the leading order of $D_{\rm up\ to\ next\ leading}$ vanish:
\begin{align}
	\label{eq:condition_determine_X}
	D_{\rm up\ to\ next\ leading}\left. \left(-\frac{5}{32}\beta^2+\frac{\epsilon^2}{6}
	+ X, \beta, \epsilon\right)\right|_{\rm leading}=0\ .
\end{align}
Thus, we can get correction terms,
\begin{align}
	\label{eq:X}
	X=\pm\frac{5}{48}\sqrt{15}\beta\epsilon^2\ .
\end{align}
Up to the next leading order, 
the particular relationships among parameters are given by
\begin{align}\label{eq:3/4_condition_D=0_next_leading}
	\chi=-\frac{5}{32}\beta^2+\frac{\epsilon^2}{6}\pm\frac{5}{48}\sqrt{15}\beta\epsilon^2\ .
\end{align}	
Remarkably,  the original curve (\ref{eq:3/4_condition_D=0}) 
splits into two curves (\ref{eq:3/4_condition_D=0_next_leading}).
In the region which intervene between (\ref{eq:3/4_condition_D=0_next_leading}),
the all order of discriminant $D_{\rm up\ to\ next\ leading}$ is negative,
{\it i.e.} $D_{\rm up\ to\ next\ leading} <0 $.
Hence, in the region between the two curves (\ref{eq:3/4_condition_D=0_next_leading}),
the cubic equation (\ref{eq:075_det_concleate_up_to_next_leading}) can be rewritten as follows:
\begin{align}
	\left[\,q+g_1(\beta, \epsilon)\,\right]\left[\,q+h_1(\beta, \epsilon)+ih_2(\beta, \epsilon)\,\right]
	\left[\,q+h_1(\beta, \epsilon)-ih_2(\beta, \epsilon)\,\right]=0\ ,
\end{align}
and the criterion for stability (\ref{eq:condition_q}) is not satisfied.
Thus, we have found that  two curves (\ref{eq:3/4_condition_D=0_next_leading})
is nothing but the transition curves for $\bar{\kappa}=3/4$.

\section{\label{sec:diagram}Discussion}
\begin{figure}[H]
	\begin{center}
		\includegraphics[width=6cm]{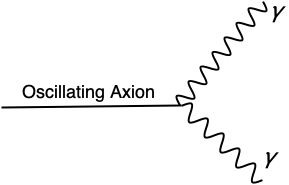}
		\caption{(\ref{eq:diagram_ordinary}) in the case of $n=1$.}
		\label{fig:diagram_ordinary}
	\end{center}
\end{figure}

In the case of the conventional Mathieu equation, bifurcation points are located at $\bar{\kappa}=n/2$.
This bifurcation point can be also rewritten as follows:
\begin{align}
	nm_a=2k\ \ \ (n=1,2,3,\cdots)\ ,\label{eq:diagram_ordinary}
\end{align}
where $m_a$ is axion mass, and $k$ is the wave number.
Diagrammatically, Eq.~(\ref{eq:diagram_ordinary}) for $n=1$ can be interpreted as in Fig.~\ref{fig:diagram_ordinary}.
Namely, the parametric resonance is nothing but a coherent decay of axions into photons.

\begin{figure}[H]
	\begin{center}
		\includegraphics[width=9cm]{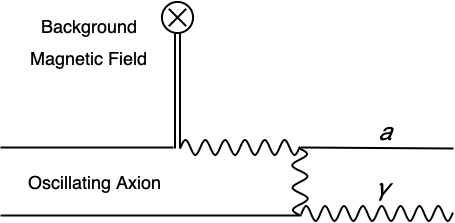}
		\caption{(\ref{eq:diagram_new}) in the case of $n=2$.}
		\label{fig:diagram_new}
	\end{center}
\end{figure}

As described in the Sec.~\ref{subsec:new} and Sec.~\ref{subsec:kappa_075},
in the situation where axion and magnetic field coexist in the background,
a new bifurcation point $\sqrt{1+\bar{\kappa}^2}+\bar{\kappa}=n$ arises.
This bifurcation point can be also rewritten as follows:
\begin{align}
	nm_a=\sqrt{m^2_a+k^2}+k\ \ \ (n\geq2,\ n=2, 3, 4, \cdots)\ .\label{eq:diagram_new}
\end{align}
The case for $n=2$ is illustrated in Fig.~\ref{fig:diagram_new}.
In this case, an axion and a photon are generated through the coherent decay of axions and photons
in the background.

Finally, let us consider what system needs to be arranged to give rise to instability 
seen in Sec~\ref{subsec:new} and Sec.~\ref{subsec:kappa_075}.
Appropriate numerical values depend on the wavelength $L$,
\begin{align}
	L \equiv \frac{2\pi}{k}\ .
\end{align}
The dimensionless parameter $\kappa$ determine a relation between the axion mass and the wavelength of electromagnetic waves as follows:
\begin{align}	
	\label{eq:kappa_microwave}
	\kappa =0.75 
	\left(\frac{1.65\times10^{-6}\,{\rm eV}}{m_a}\right)\left(\frac{10^{2}\,{\rm cm}}{L}\right) \ .
\end{align}
The parameter $\beta$ characterize the strength of magnetic fields.
In the case of a neutron star, and trying to detect the signal with microwaves
($L=10^2\,{\rm cm}$), the value of $\beta$ is given by
\begin{align}
	\label{eq:beta_microwave}
	\beta=0.12 \left(\frac{1.65\times10^{-6}\,{\rm eV}}{m_a}\right)
	\left(\frac{g_{a\gamma\gamma}}{10^{-11}\,{\rm GeV}^{-1}}\right)
	\left(\frac{B_0}{10^{15}\,{\rm G}}\right)\ .
\end{align}
The axion dark matter--photon conversion could be effective.
On the other hand, $\epsilon$ has a extremely small value 
in the case of (\ref{eq:kappa_microwave}) and (\ref{eq:beta_microwave}).
If you assume ultralight axions, $\epsilon$ has a suitable value:
\begin{align}
	\epsilon =0.097\,\left(\frac{\kappa}{0.75}\right)
	\left(\frac{g_{a\gamma\gamma}}{10^{-11}\,{\rm GeV}^{-1}}\right)
	\left(\frac{1.65\times10^{-22}\,{\rm eV}}{m_a}\right)
	\left(\frac{\rho}{0.3 \,\rm GeV/cm^3}\right)\ .
\end{align}
The de Broglie wavelength $L_{\rm dB}$ where spatial variation of axion can be neglected is as follows:
\begin{align}
	L_{\rm dB} = \frac{2\pi}{m_av}
	\sim
	0.4\,{\rm kpc}\left(\frac{10^{-22}\,{\rm eV}}{m_a}\right)\ ,
	\label{eq:de_Broglie}
\end{align}
where we took a typical velocity in the galaxy $v\sim10^{-3}$.
Since the coherence length is sufficiently long, we can expect parametric amplification of
electromagnetic waves with the wavelength $10^{18}\,{\rm cm}\sim1\,{\rm pc}$. 
Here, we would like to comment on the coherence of the axion dark matter.
We assumed coherence of the  axion dark matter,
and use the values $0.3 \,\rm GeV/cm^3$ and $v\sim10^{-3}$ 
as a rough parameter estimate.
However, for the more precise analysis, 
we need to compare the bandwidth of instability with velocity dispersion
as discussed in~\cite{10.1140/epjc/s10052-019-6759-7}.

Now, the question is whether both the conversion and the resonance can be important at the same time.
In the case of radio waves, three parameters 
included in basic equations (\ref{eq:fundamental_axion})--(\ref{eq:fundamental_perp}) are given as follows:
\begin{align}
	&\kappa =0.75 
	\left(\frac{1.65\times10^{-14}\,{\rm eV}}{m_a}\right)\left(\frac{10^{10}\,{\rm cm}}{L}\right)\ ,\\
	&\beta=0.12 \left(\frac{1.65\times10^{-14}\,{\rm eV}}{m_a}\right)
	\left(\frac{g_{a\gamma\gamma}}{10^{-11}\,{\rm GeV}^{-1}}\right)
	\left(\frac{B_0}{10^{7}\,{\rm G}}\right)\ ,\\
	&\epsilon =0.097\,\left(\frac{\kappa}{0.75}\right)
	\left(\frac{g_{a\gamma\gamma}}{10^{-11}\,{\rm GeV}^{-1}}\right)
	\left(\frac{1.65\times10^{-14}\,{\rm eV}}{m_a}\right)
	\left(\frac{\rho}{0.3\times10^{8}\,\rm GeV/cm^3}\right)\ .
\end{align}
A strong magnetic field $B_0\sim10^{7}\,{\rm G}$ can be realized with a white dwarf.
However, energy density $\rho$ needs $10^{8}$ times as much as
the average density of dark matter near the solar system.

Next, in the case of an ultralight axion, three parameters 
included in basic equations (\ref{eq:fundamental_axion})--(\ref{eq:fundamental_perp}) are given as follows:
\begin{align}
	&\kappa=0.75
\left(\frac{1.65\times10^{-22}\,{\rm eV}}{m_a}\right)\left(\frac{10^{18}\,{\rm cm}}{L}\right)\ ,\\
	&\beta=0.12\left(\frac{1.65\times10^{-22}\,{\rm eV}}{m_a}\right)
	\left(\frac{g_{a\gamma\gamma}}{10^{-11}\,{\rm GeV}^{-1}}\right)
	\left(\frac{B_0}{10^{-1}\,{\rm G}}\right)\ ,\\
	&	\epsilon =0.097\,\left(\frac{\kappa}{0.75}\right)
	\left(\frac{g_{a\gamma\gamma}}{10^{-11}\,{\rm GeV}^{-1}}\right)
	\left(\frac{1.65\times10^{-22}\,{\rm eV}}{m_a}\right)
	\left(\frac{\rho}{0.3 \,\rm GeV/cm^3}\right)\ .
\end{align}
It might be difficult to find the astrophysical situation with
the strength of magnetic fields, $B_0=10^{-1}\,{\rm G}$
and to detect electromagnetic waves $L=10^{18}\,{\rm cm}\sim1\,{\rm pc}$.

Devising a smart way,
we may be able to realize the situation where both the axion dark matter--photon conversion and
the resonance are relevant in the laboratory.

\section{\label{sec:conclusion}Conclusion}

We studied the stability of axion dark matter--photon conversion numerically and analytically.
Since the axion field is coupled with the electromagnetic field,
axions can be converted into photons and {\it vice versa}.
On the other hand, axion is one of the candidates for dark matter,
and photon propagating in axion dark matter obeys the Mathieu equations.
Therefore, it is important to understand the behavior of the system where the axion dark matter and the magnetic field coexist.
First, we derived basic equations describing axion dark matter--photon conversion.
Then, we found the instability band by numerical calculations.
Remarkably, we found the bands different from those in the conventional Mathieu equation.
Moreover, we found the shift of bifurcation point due to the magnetic fields.
More importantly, we confirmed numerical findings by using the analytical method.
In the course of the analysis, we found that the condition for the new boundary curves 
between the stability and the instability requires 
 different method from that of the conventional instability condition.
Finally, we gave graphical interpretation to the instability condition,
and comment on a possible physical application.

\acknowledgments
E. M. was in part supported by JSPS KAKENHI Grant No.~JP18J20018.
J. S. was in part supported by JSPS KAKENHI Grants No.~JP17H02894, No.~JP17K18778, 
No.~JP15H05895, No.~JP17H06359, No.~JP18H04589.
J. S. would like to thank Yukawa Institute for Theoretical Physics at Kyoto University. 
Discussions during the YITP workshop YITP-T-19-02 on "Resonant instabilities in cosmology" were useful to complete this work.
E. M. and J. S. are also supported by JSPS Bilateral Joint Research Projects (JSPS-NRF Collaboration) String Axion Cosmology. 

\appendix
\section{\label{appendix}Concrete Formulas for Analyzing Transition Curves at $\bar{\kappa}=3/4$}
In this appendix, we give concrete formulas which could not be shown in Sec.~\ref{subsec:kappa_075}.
Evaluating the determinant of $R_{3/4}(q, \chi, \beta, \epsilon)$ at the leading order,
we can get the following formula,
\begin{align}
\label{eq:075_det_concleate_leading}
&{\rm det}[R_{3/4}(q, \chi, \beta, \epsilon)]_{\rm leading}\nonumber\\
&={\frac {102515625\,}{262144}}\left[q+\frac{9}{25}\left(\chi+\frac{25}{24}\beta^2\right)^2\right]
	\left[q+\left(\chi-\frac{3}{8}\beta^2-\frac{4}{15}\epsilon^2\right)^2\right]
	\left[q+\left(\chi-\frac{4}{15}\epsilon^2\right)^2\right]\ .
\end{align}
\newpage
The determinant of $R_{3/4}(q, \chi, \beta, \epsilon)$ at the next leading order is given by
\begin{align}
\label{eq:075_det_concleate_next_leading}
&{\rm det}[R_{3/4}(q, \chi, \beta, \epsilon)]_{\rm next\ leading}\nonumber\\
&=\left( {\frac {16858125\,{\epsilon}^{2}}{131072}}+{\frac {999185625\,
\chi}{262144}}+{\frac {1004653125\,{\beta}^{2}}{1048576}} \right){q}^{3}\nonumber\\
&+ \left( -{\frac {875255625\,{\beta}^{4}{\epsilon}^{2}}{16777216}}
-{\frac {157773825\,{\epsilon}^{2}{\chi}^{2}}{131072}}
-{\frac {1278574875\,{\beta}^{2}{\epsilon}^{2}\chi}{2097152}}
-{\frac {25022925\,{\epsilon}^{4}\chi}{131072}}
+{\frac {4829625\,{\beta}^{2}{\epsilon}^{4}}{262144}}\right.\nonumber\\
&\left.+{\frac {20676718125\,{\beta}^{4}\chi}{8388608}}
+{\frac {7811690625\,{\beta}^{6}}{67108864}}
+{\frac {2239761375\,{\beta}^{2}{\chi}^{2}}{1048576}}
+{\frac {2389570875\,{\chi}^{3}}{262144}}
-{\frac {155925\,{\epsilon}^{6}}{32768}} \right) {q}^{2}\nonumber\\
&+ \left( -{\frac {55255921875\,{\beta}^{8}{\epsilon}^{2}}{1073741824}}
-{\frac {256480425\,{\epsilon}^{2}{\chi}^{4}}{131072}}
-{\frac {41518828125\,{\beta}^{10}}{1073741824}}
-{\frac {1899828675\,{\beta}^{4}{\epsilon}^{2}{\chi}^{2}}{2097152}}\right.\nonumber\\
&\left.-{\frac {1549325475\,{\beta}^{2}{\epsilon}^{2}{\chi}^{3}}{1048576}}
+{\frac {324707990625\,{\beta}^{8}\chi}{1073741824}}
+{\frac {22505596875\,{\beta}^{6}{\epsilon}^{2}\chi}{134217728}}
+{\frac {885060675\,{\beta}^{4}{\epsilon}^{4}\chi}{16777216}}\right.\nonumber\\
&\left.+{\frac {86667165\,{\beta}^{2}{\epsilon}^{4}{\chi}^{2}}{262144}}
-{\frac {154456875\,{\beta}^{6}{\epsilon}^{4}}{16777216}}
-{\frac {34266645\,{\epsilon}^{4}{\chi}^{3}}{65536}}
-{\frac {5557987125\,{\beta}^{6}{\chi}^{2}}{8388608}}\right.\nonumber\\
&\left.+{\frac {2187820125\,{\beta}^{4}{\chi}^{3}}{1048576}}
+{\frac {3551961375\,{\beta}^{2}{\chi}^{4}}{1048576}}
+{\frac {1781584875\,{\chi}^{5}}{262144}}
-{\frac {20585745\,{\beta}^{2}{\epsilon}^{6}\chi}{524288}}\right.\nonumber\\
&\left.-{\frac {19207125\,{\beta}^{4}{\epsilon}^{6}}{2097152}}
+{\frac {1971783\,{\epsilon}^{6}{\chi}^{2}}{8192}}
-{\frac {382725\,{\beta}^{2}{\epsilon}^{8}}{65536}}
-{\frac {168399\,{\epsilon}^{8}\chi}{8192}}
-{\frac {2025\,{\epsilon}^{10}}{2048}} \right) q\nonumber\\
&-{\frac {157181056875\,{\beta}^{8}{\epsilon}^{2}{\chi}^{2}}{1073741824}}
+{\frac {52579471275\,{\beta}^{6}{\chi}^{3}{\epsilon}^{2}}{134217728}}
-{\frac {2718276975\,{\beta}^{4}{\chi}^{4}{\epsilon}^{2}}{16777216}}
-{\frac {2757861675\,{\beta}^{2}{\epsilon}^{2}{\chi}^{5}}{2097152}}\nonumber\\
&-{\frac {81848475\,{\epsilon}^{2}{\chi}^{6}}{131072}}
+{\frac {324707990625\,{\beta}^{8}{\chi}^{3}}{1073741824}}
-{\frac {15516765\,{\epsilon}^{4}{\chi}^{5}}{131072}}
-{\frac {131274675\,{\beta}^{6}{\epsilon}^{4}{\chi}^{2}}{4194304}}\nonumber\\
&-{\frac {557685\,{\beta}^{2}{\epsilon}^{4}{\chi}^{4}}{32768}}
+{\frac {763069815\,{\beta}^{4}{\chi}^{3}{\epsilon}^{4}}{16777216}}
-{\frac {52275587625\,{\beta}^{6}{\chi}^{4}}{67108864}}
-{\frac {3174157125\,{\beta}^{4}{\chi}^{5}}{8388608}}\nonumber\\
&+{\frac {2316853125\,{\chi}^{6}{\beta}^{2}}{1048576}}
+{\frac {391199625\,{\chi}^{7}}{262144}}
-{\frac {41518828125\,{\beta}^{10}{\chi}^{2}}{1073741824}}
+{\frac {2560977\,{\epsilon}^{6}{\chi}^{4}}{32768}}\nonumber\\
&-{\frac {55647\,{\epsilon}^{8}{\chi}^{3}}{8192}}
+{\frac {5843390625\,{\beta}^{10}{\epsilon}^{2}\chi}{268435456}}
+{\frac {37858336875\,{\beta}^{8}{\epsilon}^{4}\chi}{1073741824}}
+{\frac {86993595\,{\beta}^{4}{\epsilon}^{6}{\chi}^{2}}{2097152}}\nonumber\\
&+{\frac {51850125\,{\beta}^{6}{\epsilon}^{6}\chi}{16777216}}
+{\frac {59875443\,{\beta}^{2}{\epsilon}^{6}{\chi}^{3}}{524288}}
-{\frac {877797\,{\beta}^{2}{\epsilon}^{8}{\chi}^{2}}{65536}}
-{\frac {4501575\,{\beta}^{4}{\epsilon}^{8}\chi}{524288}}\nonumber\\
&-{\frac {50625\,{\beta}^{4}{\epsilon}^{10}}{262144}}
-{\frac {729\,{\epsilon}^{10}{\chi}^{2}}{2048}}
-{\frac {1268915625\,{\beta}^{8}{\epsilon}^{6}}{268435456}}
-{\frac {8758125\,{\epsilon}^{8}{\beta}^{6}}{4194304}}\nonumber\\
&-{\frac {102515625\,{\beta}^{10}{\epsilon}^{4}}{33554432}}
-{\frac {18225\,{\beta}^{2}{\epsilon}^{10}\chi}{32768}}\ .
\end{align}
\newpage
A concrete expression for the leading order of discriminant $D_{\rm up\ to\ next\ leading}$
(\ref{eq:condition_determine_X}) is given by
\begin{align}
D_{\rm up\ to\ next\ leading}&\left.\left(-\frac{5}{32}\beta^2+\frac{\epsilon^2}{6}
+ X, \beta, \epsilon\right)\right|_{\rm leading}\nonumber\\
=&
{\frac {72874359645335178370006084442138671875}{158456325028528675187087900672}}\,{X}^{2}{\beta}^{
18}{\epsilon}^{2}\nonumber\\
&+{\frac {5803968607760384615992641448974609375}{19807040628566084398385987584}}\,{X}^{2}{\beta}^{16}{
\epsilon}^{4}\nonumber\\
&+{\frac {30577865593094780185718536376953125}{309485009821345068724781056}}\,{X}^{2}{\beta}^{14}{
\epsilon}^{6}\nonumber\\
&+{\frac {1437712208674487277339935302734375}{77371252455336267181195264}}\,{X}^{2}{\beta}^{12}{\epsilon}^{8}\nonumber\\
&+{\frac {4466011639849064421844482421875}{2417851639229258349412352}}\,{X}^{2}{\beta}^{10}{\epsilon}^{
10}\nonumber\\
&+{\frac {22902623794097766265869140625}{302231454903657293676544}}\,{X}^{2}{\beta}^{8}{\epsilon}^{12}\nonumber\\
&+{\frac {1514167250408630928354570865631103515625}{5070602400912917605986812821504}}\,{X}^{2}{\beta}
^{20}\nonumber\\
&-{\frac {63090302100359622014773786067962646484375
}{1298074214633706907132624082305024}}\,{\beta}^{22}{\epsilon}^{4}\nonumber\\
&-{\frac {3036431651888965765416920185089111328125}{
40564819207303340847894502572032}}\,{\beta}^{20}{\epsilon}^{6}\nonumber\\
&-{\frac {241832025323349358999693393707275390625}{
5070602400912917605986812821504}}\,{\beta}^{18}{\epsilon}^{8}\nonumber\\
&-{\frac {1274077733045615841071605682373046875}{
79228162514264337593543950336}}\,{\beta}^{16}{\epsilon}^{10}\nonumber\\
&-{\frac {59904675361436969889163970947265625}{
19807040628566084398385987584}}\,{\beta}^{14}{\epsilon}^{12}\nonumber\\
&-{\frac {186083818327044350910186767578125}{
618970019642690137449562112}}\,{\beta}^{12}{\epsilon}^{14}\nonumber\\
&-{\frac {954275991420740261077880859375}{77371252455336267181195264}}\,{\beta}^{10}{\epsilon}^{16}\ .
\end{align}

\bibliography{DK12403.bib}
\end{document}